\documentclass{vk1art}
\begin{document}
\comment{Los Alamos e-print archive: "http://arXiv.org/"}
\subject{76.80.+y, 63.20.-e}
\title{Theory of Incoherent Nuclear Inelastic\\
       Resonant Scattering of Synchrotron Radiation}
\author{{\sc V.\,G.\,Kohn\,}$^a$\footnote{e-mail: kohn@kurm.polyn.kiae.su}
        ~~and~~{\sc A.\,I.\,Chumakov\,}$^b$}
\institute{$^a$\,{\it Russian research centre ''Kurchatov Institute'',
           123182 Moscow, Russia}\\
           $^b$\,{\it European synchrotron radiation facility,
           BP-220, 38043 Grenoble, France} }
\date{(submitted 2001, November, 29)}
\abstract{
The theory of incoherent nuclear resonant scattering of synchrotron radiation
accompanied by absorption or emission of phonons in a crystal lattice is
developed. The theory is based on the Maxwell's equations and
time-dependent quantum mechanics under the condition of incoherent
scattering of radiation by various nuclei. A concept of coherence in
scattering processes, properties of the synchrotron radiation, and
conditions of measurement are discussed. We show that employing the
monochromator with a narrow bandwidth plays a decisive role. The
equations for energy dependence and time evolution of scattered radiation
are derived in detail for the case of the non-split nuclear levels. The
effect of hyperfine interaction on the time spectra is also considered.
We show that the Singwi and Sj\"{o}lander theory of inelastic nuclear
absorption, improved by the convolution with the instrumental function of
the monochromator, is valid for the incoherent nuclear scattering in the
case of integral over time measurements.
}
\maketitle

\section*{1. Introduction}

Nuclear resonant scattering of x rays attracts much interest after the
M\"{o}ssbauer's discovery of pure elastic nuclear resonant scattering
\cite {Mo}. The recoil energy is transferred to an entire crystal lattice,
and the frequency of the scattered by nucleus radiation remains resonant.
This gives an access to studies of coherent phenomena in nuclear
scattering and stimulates a development of theory of coherent elastic
nuclear resonant scattering (see the review of theoretical results in
Refs. [2,3].
It became clear soon after the discovery that,
because of the very small energy width of nuclear resonance, nuclear
resonant scattering gives a good opportunity for studying dynamic of
nuclear motions in crystals, especially thermal vibration and diffusion
[4--6].
 However, measurements of phonon spectra with
radioactive sources were extremely difficult due to a small cross-section
of inelastic processes.

Recently the situation has drastically changed owing to the development
of the synchrotron radiation sources of third generation such as ESRF
(France), APS (USA) and SPring8 (Japan). A high brightness and spectral
density of these sources, on the one hand, and a development of x-ray
monochromator with energy bandwidth less than one millielectronvolt,
on the other hand, give rise to numerous experimental studies of nuclear
resonant inelastic absorption. These experiments deal with measurements
of atomic fluorescence, which results from resonant absorption of
incident photon by nucleus with simultaneous absorption or emission of
phonons in solids. The review of experimental works and details of
experimental methods are discussed in Ref. \cite{CS}. The analysis of the
experimental results in the first works is based on the theory
by Singwi and Sj\"{o}lander \cite{SS}, which is, strictly speaking,
formulated only for the particular case of cubic Bravais crystal lattice.
The general theory of nuclear inelastic absorption for anisotropic
crystals is developed in Ref. \cite{KCR}.

We note that the scattering process is not discussed in these works.
Only that part of nuclear absorption cross-section, which corresponds to
absorption of photon by thermally vibrating nucleus is
considered. The incident radiation is assumed as a monochromatic
wave, i.e., the specific properties of the synchrotron radiation, the
details of measurement and the time dependence of scattered radiation are
not considered. One can believe that the theory is approximately
valid for the nuclear resonant inelastic absorption accompanied by the
atomic fluorescence, however, being not adapted to the real experimental
conditions. As for the theory of nuclear resonant inelastic scattering
(nuclear fluorescence) the problem remained open before recently. An
attempt of developing the theory of both the nuclear fluorescence and
the atomic fluorescence, which follow the process of inelastic nuclear
excitation, is presented in Ref. \cite{SK} using methods of quantum
electrodynamics and Green functions. However, in our opinion, the theory
is not complete in the part where the details of experiment are discussed.
Also the complicated technique does not allow obvious description of
the scattering processes. The theory is difficult for understanding and
requires special theoretical background.

This limitation can be overcome in frames of another approach which,
after the Laue's \cite{La} work, is widely used for description of
various scattering processes. It is based on the Maxwell's equations and
dielectic properties of medium. In many works (see, for example, Ref.
\cite{Ka}) nuclear resonant scattering is described by Maxwell's equation
where a source of scattered radiation is treated as nuclear currents
calculated by means of the time dependent perturbation theory of
quantum mechanics \cite{He}. In this work we present the theory of
incoherent nuclear inelastic resonant scattering within this
conventional approach. This allows us to take explicitly into account
properties of the synchrotron radiation  and conditions of
measurements.

\section*{2. Maxwell's equation and coherence}

For x rays with the energy about 10 keV and more, the electric
$\mathbf{E}(\mathbf{r},t)$ and magnetic $\mathbf{B}(\mathbf{r},t)$ fields
are strictly connected to each other, and the values of the modulo
squared field strength are identical after averaging over a period of
oscillations . Therefore it is sufficient to consider an equation for
any of these fields, or for some vector which allows one to determine
both of them. In the quantum theory of radiation (see, for example,
\cite{Sa}) the vector potential $\mathbf{A}(\mathbf{r},t)$ in the
Coulomb gauge is used. It is chosen in such a way that
\begin{equation}
 \mathbf{\nabla \cdot A}=0\,,\quad \mathbf{B}=\mathbf{\nabla \times A}\,,
 \quad\mathbf{E}=-\frac{1}{c}\frac{\partial \mathbf{A}}{\partial t}\,.
 \label{ABE}
\end{equation}
 The Maxwell's equation for the vector potential can be written as
\begin{equation}
 \mathbf{\nabla}^{2}\mathbf{A(\mathbf{r}},t)-\frac{1}{c^{2}}
 \frac{\partial^{2}}{\partial t^{2}}\mathbf{A}(\mathbf{r},t)=-
 \frac{1}{c}\,\mathbf{j}(\mathbf{r},t)\,,  \label{AAj}
\end{equation}
where $\mathbf{j}(\mathbf{r},t)$ is the current acting as a source of
electromagnetic radiation, $c$ is the speed of light.

\subsection*{2.1 Coherence as a possibility to observe interference.}

The current $\mathbf{j}(\mathbf{r},t)$ results from moving charged
particles. If the particles move independently of the radiation, they are
regarded as free sources of radiation. When the motion of particles is
forced by incident radiation, we deal with re-radiation or scattering. In
the latter case we deal with the induced current, which is proportional
to the radiation field strength. The Maxwell's equation is linear
relative to the field strength, and the induced currents for the x rays
are also linear with high accuracy. Therefore the equation obeys the
superposition principle for any ensemble of sources and scatterers. In
other words, any fields and any scattering processes are coherent to each
other.

This well known fact of classical optics should be stressed here, because
sometimes one speaks about a principal incoherence of some sources or
scattering processes. In reality, the problem of coherence absence is
connected with conditions of measurement. Though the intensity of
radiation from various sources or scatterers is always positive and
must be only added, the interference term in the modulo squared total
field strength has alternating signs in various time-space regions. As a
rule, it strongly oscillates in time and space and its average value
vanishes. Therefore, if experimental conditions allows only a measurement
of average value of the interference term, the results cannot show
coherent phenomena. However, if an experimental technique and properties
of incident radiation allows one to observe the interference term, the
processes are coherent. Thus, in order to observe the coherence one needs
specific experimental conditions with essentially high spatial or
temporal resolution of the detector.

For example, consider excitation of nuclei by x rays with extremely
broad energy range containing all nuclear resonances. In this case,
nuclear resonant scattering via various nuclear levels creates waves with
so large difference in frequencies, that it would be impossible to
record the interference term even with the best modern technique.
Therefore, these acts of scattering should be considered as incoherent
\cite{SK}. However, scattering via various sub-levels of hyperfine
structure of a single nuclear level leads to quantum beats in the
time spectra with the period of about tens nanoseconds. Time
resolution of available detectors allows measurements of resulting
interference (see the review of experimental results in Ref. \cite{UB}).
Thus, these quantum beats is a result of coherent phenomena.

Similar situations take place in space spectra. For example, if a source
is strongly localized and a detector has sufficiently high spatial
resolution, one can expect to record the interference phenomena even with
x rays of high energy \cite{KSS}. In general, the concept of coherence
can be formulated as a possibility to observe the interference term in
the intensity of radiation recorded from many sources and/or scatterers.

There exist particular conditions, when electric fields, resulted
from scattering of primary radiation by very many nuclei, are added with
the identical phase factor. Then the total intensity of scattered wave
is proportional to $N^2$ where $N$ is a number of nuclei. In this case
the interference term can significantly exceed the average intensity, and
the strength of the scattered wave can be comparable with the strength of
the incident wave. The effective wave field, which excites the nuclei, is
not known a priory in this case, and it should be calculated
self-consistently. These conditions are called dynamical or multiple
scattering. A conventional approach of dynamical theory of scattering
supposes a calculation of the field inside the volume of scattering
medium solving the self-consistent equations. The medium is considered as
infinite or limited in one direction (lamina-like medium). The fields at
both sides of the entrance surface of the sample must be identical
(boundary conditions). The field at the exit surface is considered
usually as that measured by a detector, although successive propagation
of the wave in air may be important in some cases. In the dynamical
theory, it is sometimes convenient to introduce a continuous spatial
distribution of the scatterers and the medium polarization similarly
to classical optics. A particular case of the dynamical theory is
the kinematical theory, where the scattered wave remains much weaker than
the incident wave, but scattering is still spatially coherent and
depends on the shape of the sample.

Kinematical scattering is often identified with the first Born
approximation although this is not completely correct. In the first Born
approximation the boundary problem is not considered at all. The
scatterers are assumed to be far away from both the source and
the detector, whereas the scattering amplitudes are expected to be very
small. If the experimental conditions allow one to record the interference
term, the scattering process is coherent. If the interference term can be
expected as negligible, it is sufficient to calculate the intensity of
scattered wave from one nucleus. A total intensity of radiation becomes
proportional to a number of nuclei in a target. In this approach,
however, the act of scattering should not necessarily be connected to a
single nucleus. In general, the incident wave illuminates many nuclei.
The problem of localization or de-localization of excitation is, in fact,
identical to the problem of phasing or de-phasing of the interference
term in the total intensity of the radiation scattered by many nuclei.

In this work, we consider nuclear resonant scattering accompanied by
absorption or emission of phonons in a target. The process has
relatively small cross-section. Therefore scattering events are collected
within a large solid angle and are accumulated after many flashes of
incident synchrotron radiation. These conditions do not allow us to
record the interference between various nuclei. Thus, we can restrict
ourselves by an analysis of the incoherent scattering. We will use the
first Born approximation in the problem of scattering by a single nucleus.
A summation over many nuclei leads to a trivial factor of a number of
nuclei in the unit volume of the target. The task is solved in three
independent steps. Firstly, we declare the properties of the incident
radiation $\mathbf{A}_{i}%
\mathbf{(\mathbf{r}},t)=\mathbf{e}_{i}A_{i}\mathbf{(\mathbf{r}},t)$.
Secondly, we calculate the current
$\mathbf{j}_{fi}(\mathbf{r},t)$, of nuclear resonant
transitions. Finally, we calculate the scattered radiation
$\mathbf{A}_{f}\mathbf{(\mathbf{r}},t)=\mathbf{e}_{f}A_{f}
\mathbf{(\mathbf{r}},t)$, at large distance from the nucleus.
Here $\mathbf{e}_{i}$ è $\mathbf{e}_{f}$ are the unit vectors of
polarizations.

\subsection*{2.2 Temporal and spatial structure of incident wave.}

Properties of synchrotron undulator radiation are well known. They are
complicated by significant relativistic phenomena. Instead of using
them straightforwardly, we consider a simplified model, where the
current of the source is strictly localized in time and space.
\begin{equation}
\mathbf{\nabla }^{2}A_{i}\mathbf{(\mathbf{r}},t)-\frac{1}{c^{2}}\frac{%
\partial ^{2}}{\partial t^{2}}A_{i}(\mathbf{r},t)=-\frac{1}{c}\,\delta (%
\mathbf{r}-\mathbf{r}_{0})\delta (t-t_{s})\,. \label{eqAi}
\end{equation}
Here $\mathbf{r}_{0}$ is the source position, and $t_{s}$ is the instant
of flash. Although a bunch of electrons in a storage ring consists of
many electrons, the radiation from various electrons can be considered as
incoherent. Therefore each event of nuclear scattering is initiated by
incident radiation from a single electron. The width of the synchrotron
radiation pulse (e.g., about 100 ps at the ESRF, Grenoble) does not enter
the problem.

The solution of eq. (\ref{eqAi}) is obtained as the Fourier
integral
\begin{equation}
 A_{i}(\mathbf{r},t)=\int \frac{d\omega}{2\pi }\exp (-i\omega \lbrack
 t-t_{s}])\frac{\exp (i\omega |\mathbf{r}-\mathbf{r}_{0}|/c)}
 {4\pi c|\mathbf{r}-\mathbf{r}_{0}|}\,,  \label{Ai1}
\end{equation}
where the particular solution for the monochromatic source is applied
as a spherical wave. The integral in (\ref{Ai1}) can be calculated
analytically as follows
\begin{equation}
A_{i}(\mathbf{r},t)=\frac{\delta (t-t_{s}-|\mathbf{r}-\mathbf{r}_{0}|/c)}{%
4\pi c|\mathbf{r}-\mathbf{r}_{0}|}\,.  \label{s3}
\end{equation}
This result has a simple physical meaning. Instantaneous localized
excitation spreads in all directions from the point $\mathbf{r}_{0}$\,
and is distributed over the surface of the sphere of radius
$|\mathbf{r}-\mathbf{r}_{0}|$. At the instant $t$ the radius is equal to
the distance $|\mathbf{r}-\mathbf{r}_{0}|=c(t-t_{s})$ which the light
covers during the time interval $(t-t_{s})$. The intensity of the
radiation is proportional to $|\mathbf{r}-\mathbf{r}_{0}|^{-2}$.
The total intensity over the sphere area remains constant.

In further calculations, the integral presentation (\ref{Ai1}) is more
convenient. At large distances from the source
$|\mathbf{r}_{0}|\gg |\mathbf{r}|$ we can use the approximation
\begin{equation}
A_{i}(\mathbf{r},t)\approx \frac{1}{4\pi c|\mathbf{r}_{0}|}\int \frac{%
d\omega }{2\pi }\exp (i\mathbf{k}_{0}^{\prime }\mathbf{r}-i\omega \lbrack
t-t_{s}-t_{0}])\,,  \label{Ai2}
\end{equation}
where $t_{0}=|\mathbf{r}_{0}|/c$ is the flight time through the distance
$|\mathbf{r}_{0}|$ from the source
to the origin of the our coordinate system where the nucleus is located,
$\mathbf{k}_{0}^{\prime }=(\omega /c)\mathbf{s}_{0}$,
$\mathbf{s}_{0}=-\mathbf{r}_{0}/|\mathbf{r}_{0}|$. We use the
approximation $|\mathbf{r}-\mathbf{r}_{0}|\approx|\mathbf{r}_{0}|+
\mathbf{s}_{0}\mathbf{r}$. The expression (\ref{Ai2})
describes the set of plane monochromatic waves with all possible
frequencies.

In reality, an electron in the storage ring
is not localized in space but moves with the near-light speed.
Therefore, the synchrotron radiation is strongly anisotropic. It is
close to the spherical wave only in the electron coordinate system
\cite{Te,Ma}. At the point far from the source and in the laboratory
coordinate system, it can be approximated by the Gauss beam directed
along the speed of motion. Within its central part, however,
the radiation wave field is close to the spherical wave which
approximately coincides with the plane wave. Thus, the postulated
spatial localization is consistent with experimental conditions.
In order to estimate the temporal localization of single electron
radiation flash, we can consider an effective time width as given by
frequency spectrum of synchrotron radiation under typical experimental
conditions and the uncertainty principle. The estimated time width
turns out to be negligibly small compared to the nuclear life time. Thus,
eq.(\ref{Ai1}) approximately describes the incident synchrotron radiation
under the typical experimental conditions.

In experiments on nuclear inelastic scattering the bandwidth of
radiation is limited by a monochromator. Let the transmission amplitude
of the monochromator be $P(\omega -\omega _{0})$, where $\omega _{0}$
is the frequency at the maximum of transmissivity through the
monochromatoron, while the instrumental function 
$I_{M}(\omega )=\left| P(\omega )\right| ^{2}$ is centered at zero
argument. The frequency band,
provided by the monochromator, is sufficiently small to neglect the
frequency dependence of the wave vector
$\mathbf{k}_{0}^{\prime }$, because the considered values of the vector
$|\mathbf{r}|$ are small and comparable with displacements due to
the thermal motion of the nucleus. As a result, we arrive to the following
approximation for the amplitude of the incident radiation
\begin{equation}
A_{i}(\mathbf{r},t)\approx \frac{\exp (i\mathbf{k}_{0}\mathbf{r}-i\omega
_{0}[t-t_{s}-t_{0}])}{4\pi c|\mathbf{r}_{0}|}\int \frac{d\omega }{2\pi }
\exp(-i\omega \lbrack t-t_{s}-t_{0}])P(\omega )\,.  \label{Ai3}
\end{equation}
Hereafter $\mathbf{k}_{0}=(\omega _{0}/c)\mathbf{s}_{0}$.

The frequency integral in eq.(\ref{Ai3}) describes time evolution
of the incident radiation flash after passing through
the monochromator. It is determined by the response function
of the monochromator
\begin{equation}
\tilde{P}(t)=\int \frac{d\omega }{2\pi }\exp (-i\omega t)P(\omega )\,.
\label{s6}
\end{equation}
According to general physical principles, $P(\omega )$ as a
function of complex variable $\omega $ can have the poles only in the
bottom half-plane. Therefore the response function $\tilde{P}(t)$
vanishes for $t<0$. In other words, it describes the retarded
response. The nucleus perceives the radiation flush at the instant
$t=t_{s}+t_{0},$ which is by $t_{0}$ later
than $t_{s}$ -- the instant of the flash at the source . Time
$t_{0}$ is required for light to travel the distance $|\mathbf{r}_{0}|$.
The accurate time profile of the flash perceived by the nucleus
is determined by the monochromator and is described by the function
$\tilde{P}(t)$.

\subsection*{2.3 Temporal and spatial structure of scattered wave.}

It will be shown below that the current of nuclear transition can be
expressed as an integral over a limited range of frequencies near the
middle frequency $\omega _{1}$,
\begin{equation}
\mathbf{e}_{f}\,\mathbf{j}_{fi}(\mathbf{r},t)=\int \frac{d\omega }{2\pi }%
\exp [-i(\omega _{1}+\omega )t]J(\mathbf{r},\omega )\,.  \label{sr1}
\end{equation}
Therefore we can write the amplitude of the scattered wave at
the detector position also as a spherical monochromatic waves:
\begin{equation}
A_{f}(\mathbf{r}_{1},t_{d})=\int \frac{d\omega }{2\pi }\exp [-i(\omega
_{1}+\omega )t_{d}]\int d\mathbf{r}\frac{\exp (ik_{1}^{\prime }|\mathbf{r}%
_{1}-\mathbf{r}|)}{4\pi c|\mathbf{r}_{1}-\mathbf{r}|}J(\mathbf{r},\omega )\,,
\label{b3}
\end{equation}
where $t_{d}$\ is the instant of arriving the radiation flash to the
detector, and $k_{1}^{\prime }=(\omega_{1}+\omega )/c$.

We assume that the distance $|\mathbf{r}_{1}\mathbf{|}$\ between the
detector and the scatterer is much larger than the typical displacement
of the nucleus $|\mathbf{r}|$ due to thermal vibrations. We place again
the centre of the scatterer at the origin of the coordinate system
and use the relation $%
\mathbf{r}_{1}-\mathbf{r}=|\mathbf{r}_{1}\mathbf{|}\mathbf{s}_{1}-
\mathbf{r}$, where $\mathbf{s}_{1}\mathbf{=r}_{1}/|\mathbf{r}_{1}|$.
Then, applying an approximation $|\mathbf{r}_{1}-\mathbf{r}|
\approx |\mathbf{r}_{1}\mathbf{|-s}_{1}\mathbf{r,}$ we arrive to
\begin{equation}
A_{f}(\mathbf{r}_{1},t_{d})=\int \frac{d\omega }{2\pi }\frac{\exp [-i(\omega
_{1}+\omega )(t_{d}-t_{1})]}{4\pi c|\mathbf{r}_{1}|}\int d\mathbf{r}\exp (-i%
\mathbf{k}_{1}\mathbf{r})J(\mathbf{r},\omega )\,,  \label{sr3}
\end{equation}
Here $\mathbf{k}_{1}$ is the scattering vector,
$\mathbf{k}_{1}=k_{1}\mathbf{s}_{1}=k_{1}\mathbf{r}_{1}/|\mathbf{r}_{1}|$.
Its modulus is $k_{1}=\omega _{1}/c$, because we can neglect
small frequency $\omega $ due to the same reason
as for the incident radiation. The value $t_{1}=|\mathbf{r}_{1}|/c$
is the time interval required for light to travel the
distance $|\mathbf{r}_{1}|$ from the nucleus to the detector.

Thus, the scattered radiation from single nucleus is described by an
anisotropic spherical wave, which shape is determined by properties of
the nucleus.

\section*{3. Quantum current of nuclear transitions.}

The scatterer is the nucleus located in the node of crystal lattice of
the target. The state of scatterer can be described in terms of the
collective state of the centres of mass of all nuclei in the crystal
$|\chi _{i}\rangle $ and the internal states of the nucleus
$|\phi _{i}\rangle $. Consider first the internal nuclear
transitions.

\subsection*{3.1 The Schr\"{o}dinger equation.}

The current $\mathbf{j}(\mathbf{r},t)$ is the quantum mechanical average
value of the current density operator over the states in presence of
the electromagnetic field of radiation. Since the notation
$|\phi _{j}\rangle $ is used for the stationary states of the nucleus
we will denote the wave function in the presence of the radiation field
by $|\psi _{j}(t)\rangle $.

We consider a 3-level system with $|\phi _{i}\rangle f_{i}(t)$ as
the initial state, $|\phi _{j}\rangle f_{j}(t)$ as the intermediate
state and $|\phi_{f}\rangle f_{f}(t)$ as the final state of the nucleus.
The equation for the quasi-stationary states is
\begin{equation}
\hat{H}_{0}|\phi _{j}\rangle =E_{j}|\phi _{j}\rangle \,,\quad
f_{j}(t)=\exp (-i\omega _{j}t-\gamma _{j}t)\,,\quad \omega _{j}=
E_{j}/\hbar\,,\quad \gamma _{j}=\Gamma _{j}/2\hbar \,.  \label{a1}
\end{equation}
Here $\hat{H}_{0}$ is the Hamiltonian of the nucleus. We explicitly
take into account a decay of the excited states with time. For the
sake of simplicity, in this section we assume, that the hyperfine
splitting of the nuclear levels is absent. In the problem of the energy
spectrum of inelastic nuclear scattering with participation of phonons
the hyperfine splitting is not essential at all. It influences only the
temporal characteristics of scattering. This question will be
analyzed later.

The ground state of the nucleus $|\psi _{i}(t)\rangle $
is perturbed by the field of incident radiation
$\mathbf{A}_{i}\mathbf{(\mathbf{r}},t\mathbf{)}$. The
Hamiltonian $\hat{H}_{int}(t)$ of interaction of the nucleus with the
radiation field is small and can be taken into account in frames of
perturbation theory. In general, the perturbed ground state contains
small contributions from other nuclear states:
\begin{equation}
|\psi _{i}(t)\rangle =|\phi _{i}\rangle f_{i}(t)+a_{ji}(t)|\phi _{j}
\rangle f_{j}(t)+\cdots \,.  \label{a2}
\end{equation}
This function is the solution of the Schr\"{o}dinger equation
\begin{equation}
i\hbar \frac{\partial \psi _{i}(t)}{\partial t}=[\hat{H}_{0}+\hat{H}%
_{int}(t)]\psi _{i}(t)\,.  \label{a3}
\end{equation}
A substitution of (\ref{a2}) into (\ref{a3}) leads to
\begin{equation}
i\hbar \frac{\partial f_{i}}{\partial t}\phi _{i}+a_{ji}i\hbar \frac{%
\partial f_{j}}{\partial t}\phi _{j}+i\hbar \frac{\partial a_{ji}}{\partial t%
}\phi _{j}f_{j}\approx \hat{H}_{0}\phi _{i}f_{i}+a_{ji}\hat{H}%
_{0}\phi _{j}f_{j}+\hat{H}_{int}(t)\phi _{i}f_{i}\,.  \label{a4}
\end{equation}
Since the first and second terms from both sides are equal to each other,
we obtain
\begin{equation}
i\hbar \frac{\partial a_{ji}}{\partial t}=\langle \phi _{j}|\hat{H}%
_{int}(t)|\phi _{i}\rangle f_{i}(t)f_{j}^{-1}(t)\,.  \label{a5}
\end{equation}
Assuming the adiabatic inclusion of the interaction, i.e.
$\hat{H}_{int}(-\infty )=0$, we arrive to expression
\begin{equation}
a_{ji}(t)=\frac{1}{i\hbar }\int_{-\infty }^{t}dt^{\prime }\exp (i\omega
_{ji}t^{\prime }+\gamma _{j}t^{\prime })\langle \phi _{j}|\hat{H}%
_{int}(t^{\prime })|\phi _{i}\rangle \,.  \label{a6}
\end{equation}
Here $\omega _{ji}=\omega _{j}-\omega _{i}$. We explicitly took into
account that $\gamma _{i}=0$.

\subsection*{3.2 Induced current as a source of scattered radiation.}

Let us consider the matrix element of the current density operator for
the transition of the second order from the initial state $i$ into the
state $f$ assuming that the transition is realized through the
intermediate state $j$ which is excited by a radiation transition from
$i$ into $j$. The final state $f$ must be considered with zero
radiative width, because a finite width describes transition probability
and, therefore, has sense only for the intermediate state. Applying the
Fourier transformation we have
\begin{equation}
\mathbf{j}_{fi}(\mathbf{r},t)=\int \frac{d\mathbf{k}}{(2\pi )^{3}}e^{i%
\mathbf{k}[\mathbf{r}-\mathbf{r}_{n}(t)]}e^{i\omega _{f}t}\langle \phi
_{f}\,|\,\hat{\mathbf{j}}_{n}(\mathbf{k})|\psi _{i}(t)\rangle \,,
\label{n4}
\end{equation}
where $\mathbf{r}_{n}(t)$ is the coordinate of the vibrated nucleus.
Resonant scattering is described by the perturbed term in (\ref{a2}).
In accord with (\ref{a6}) we have
\begin{eqnarray}
\mathbf{j}_{fi}(\mathbf{r},t) &=&\frac{\exp (i\omega _{fi}t)}{i\hbar }\int
\frac{d\mathbf{k}}{(2\pi )^{3}}\int_{-\infty }^{t}dt^{\prime }e^{i\mathbf{k[r%
}-\mathbf{r}_{n}(t)]}\langle \phi _{f}\/|\,\hat{\mathbf{j}}_{n}(%
\mathbf{k})|\phi _{j}\rangle  \nonumber \\
&&\times \langle \phi _{j}\/|\,\hat{H}_{int}(t^{\prime })|\phi
_{i}\rangle \exp [i(-\omega _{ji}+i\gamma _{j})(t-t^{\prime })]\,.  \label{n7}
\end{eqnarray}

The Hamiltonian of interaction is well known. It is convenient to write
it in the form
\begin{eqnarray}
\hat{H}_{int}(t^{\prime }) &=&-\frac{1}{c}\int d\mathbf{r}^{\prime }%
\,\hat{\mathbf{j}}_{n}(\mathbf{r}^{\prime }\mathbf{-r}_{n})%
\mathbf{A}_{i}(\mathbf{r}^{\prime },t^{\prime })  \nonumber \\
&=&-\frac{1}{c}\int d\mathbf{r}^{\prime }\int \frac{d\mathbf{k}^{\prime }}{%
(2\pi )^{3}}e^{i\mathbf{k}^{\prime }(\mathbf{r}^{\prime }-\mathbf{r}%
_{n}(t^{\prime }))}\mathbf{e}_{i}\hat{\mathbf{j}}_{n}(\mathbf{k}^{\prime
})\,A_{i}(\mathbf{r}^{\prime },t^{\prime })\,,  \label{n8}
\end{eqnarray}
where, as above, $\mathbf{e}_{i}$ is the unit vector of polarization
of incident radiation. Then, the projection of the current
density on the unit vector of polarization of the scattered wave may be
written in the form
\begin{equation}
\mathbf{e}_{f}\,\mathbf{j}_{fi}(\mathbf{r},t)=\int_{-\infty }^{t}dt^{\prime
}\int d\mathbf{r}^{\prime }\,M_{fi}(\mathbf{r},t;\mathbf{r}^{\prime
},t^{\prime })A_{i}(\mathbf{r}^{\prime },t^{\prime })\,,  \label{efjfi}
\end{equation}
where we introduce the scattering matrix
\begin{eqnarray}
M_{fi}(\mathbf{r},t;\mathbf{r}^{\prime },t^{\prime }) &=&\frac{i\exp
(i\omega _{fi}^{(n)}t)}{\hbar c}\int \frac{d\mathbf{k}d\mathbf{k}^{\prime }}{%
(2\pi )^{6}}e^{i\mathbf{kr+}i\mathbf{k}^{\prime }\mathbf{r}^{\prime }}N_{fi}(%
\mathbf{k},\mathbf{k}^{\prime })  \nonumber \\
&&\times \exp [i(-\omega _{ji}^{(n)}+i\gamma _{j})(t-t^{\prime })]e^{-i%
\mathbf{kr}_{n}(t)}e^{-i\mathbf{k}^{\prime }\mathbf{r}_{n}(t^{\prime })}\,.
\label{n10}
\end{eqnarray}
Hereafter the frequencies $\omega _{fi}^{(n)}$ related to the
nuclear transitions are marked by the index $(n)$ and we use notation
\begin{equation}
N_{fi}(\mathbf{k},\mathbf{k}^{\prime })=\langle \phi _{f}\/|\,\mathbf{%
e}_{f}\hat{\mathbf{j}}_{n}(\mathbf{k})|\phi _{j}\rangle \langle \phi
_{j}\/|\,\mathbf{e}_{i}\hat{\mathbf{j}}_{n}(\mathbf{k}^{\prime
})|\phi _{i}\rangle \,.  \label{N1}
\end{equation}
Expression (\ref{n10}) for the scattering matrix contains the
coordinates of the nucleus at the instant of excitation
$t^{\prime }$ and at the instant of relaxation $t$.

The nuclear coordinates depend on the time owing to thermal vibrations.
As known, the thermal vibrations of atoms in a crystal lattice exhibit
distinct quantum behaviour, thus they must be described as absorption
and emission of phonons. Therefore we will consider the coordinate
$\mathbf{r}_{n}$ as the operator and introduce the wave function
of phonons.
\begin{eqnarray}
e^{-i\mathbf{kr}_{n}(t)}e^{-i\mathbf{k}^{\prime }\mathbf{r}_{n}(t^{\prime
})}&=&\sum_{m}\langle \chi _{f}\,|e^{-i\mathbf{kr}_{n}}|\chi _{m}\/\rangle
\langle \chi _{m}\,|e^{-i\mathbf{k}^{\prime }\mathbf{r}_{n}}|\chi
_{i}\/\rangle \exp (i\omega _{fm}^{(p)}t+i\omega _{mi}^{(p)}t^{\prime })
\nonumber \\
&=&\exp (i\omega _{fi}^{(p)}t)L_{fi}(\mathbf{k},\mathbf{k}^{\prime
},t-t^{\prime })\,.  \label{a8}
\end{eqnarray}
Here index $(p)$ shows that the frequency is related to the phonon
system. The sum over intermediate states of the phonon system is
required in order to describe an evolution of the system from the
instant $t^{\prime }$ into the instant $t$. The function $L_{fi}$
depends only on the time interval and is defined
by expression
\begin{equation}
L_{fi}(\mathbf{k},\mathbf{k}^{\prime },t)=\sum_{m}\langle \chi _{f}\,|e^{-i%
\mathbf{kr}_{n}}|\chi _{m}\/\rangle \langle \chi _{m}\,|e^{-i\mathbf{k}%
^{\prime }\mathbf{r}_{n}}|\chi _{i}\/\rangle \exp (-i\omega _{mi}^{(p)}t)\,.
\label{a9}
\end{equation}

As a result, the scattering matrix takes the form
\begin{eqnarray}
M_{fi}(\mathbf{r},t;\mathbf{r}^{\prime },t^{\prime }) &=&\frac{i\exp
(i\omega _{fi}t)}{\hbar c}\exp [i(-\omega _{r}+i\gamma )(t-t^{\prime })]
\nonumber \\
&&\times \int \frac{d\mathbf{k}d\mathbf{k}^{\prime }}{(2\pi )^{6}}e^{i%
\mathbf{kr+}i\mathbf{k}^{\prime }\mathbf{r}^{\prime }}N_{fi}(\mathbf{k},%
\mathbf{k}^{\prime })L_{fi}(\mathbf{k},\mathbf{k}^{\prime },t-t^{\prime })\,,
\label{Mt1}
\end{eqnarray}
where we introduce the total transition frequency
$\omega _{fi}=\omega _{fi}^{(n)}+\omega _{fi}^{(p)}$,
the resonance frequency $\omega _{r}=\omega _{ji}^{(n)}$
and omit the index of the width of the excited state
$\gamma =\gamma _{j}$. Expression (\ref{Mt1}) is close to
eq.(25) of Ref. \cite{SK}. However, here it is derived for 3-level
system within standard quantum mechanics approach.
A summation over the intermediate states of the nucleus, as made
in Ref. \cite{SK}, does not have sense because the resonance can be
realized only for one transition.

Substituting eqs. (\ref{Mt1}) and (\ref{Ai3}) into eq.
(\ref{efjfi}) and making identical transformations we obtain
the explicit form for the function $J(\mathbf{r},\omega )$ in eq.
(\ref{sr1}) as follows
\begin{eqnarray}
J(\mathbf{r},\omega ) &=&P(\omega )\frac{i\exp [i(\omega _{0}+\omega
)(t_{s}+t_{0})]}{4\pi \hbar c^{2}|\mathbf{r}_{0}|}\int \frac{d\mathbf{k}}{%
(2\pi )^{3}}\exp (i\mathbf{kr})\,N_{fi}(\mathbf{k},-\mathbf{k}_{0})
\nonumber \\
&&\times \tilde{L}_{fi}(\mathbf{k},-\mathbf{k}_{0},\omega _{0}+\omega
-\omega _{r}+i\gamma )  \label{n12}
\end{eqnarray}
Here $\omega _{1}=\omega _{0}-\omega _{fi}$ is the middle frequency of
the scattered wave. We also introduced the function
\begin{equation}
\tilde{L}_{fi}(\mathbf{k},\mathbf{k}^{\prime },\omega )=\int_{0}^{\infty
}dt\exp (i\omega t)L_{fi}(\mathbf{k},\mathbf{k}^{\prime },t)  \label{lpi}\,.
\end{equation}

\section*{4. Properties of scattered radiation.}

Substituting (\ref{n12}) in (\ref{sr3}) we obtain the vector potential
of the scattered wave as follows
\begin{eqnarray}
A_{f}(\mathbf{r}_{1},t_{d}) &=&\frac{\exp [-i\omega _{1}(t_{d}-t_{1})]\exp
[i\omega _{0}(t_{s}+t_{0})]}{(4\pi )^{2}c^{3}\hbar |\mathbf{r}_{0}||\mathbf{r%
}_{1}|}N_{fi}(\mathbf{k}_{1},-\mathbf{k}_{0})  \nonumber \\
&&\times \int \frac{d\omega }{2\pi }\exp (-i\omega t)\tilde{L}_{fi}(%
\mathbf{k}_{1},-\mathbf{k}_{0},\omega _{0}+\omega -\omega _{r}+i\gamma
)P(\omega )\,.  \label{sr4}
\end{eqnarray}
Here $t=t_{d}-t_{s}-t_{0}-t_{1}$ is the time delay of radiation in the
monochromator and in the nucleus, because $t_{d}$ is the time instant
of arrival of the scattered flash to the detector, whereas
$(t_{s}+t_{0}+t_{1})$ have sense of the time instant of arrival of the
flash propagating in vacuum along the identical trajectory.
The definition (\ref{lpi}) states that the function
$\tilde{L}_{fi}(\mathbf{k}_{1},-\mathbf{k}_{0},\omega )$ has the poles
only in the bottom half-plane of the complex variable $\omega $.
Therefore the strength of the scattered radiation as function of the
delay time $t$ vanishes for $t<0$.
As a result, we obtain general expression for intensity
of scattered radiation as a function of the delay time $t$ and
the mean frequency $\omega _{0}$ of the monochromator transmissivity as
\begin{equation}
I(\omega _{0},t)=I_{0}I_{N}\left| \int \frac{d\omega }{2\pi }\exp (-i\omega
t)\tilde{L}_{fi}(\mathbf{k}_{1},-\mathbf{k}_{0},\omega _{0}+\omega
-\omega _{r}+i\gamma )P(\omega )\right| ^{2}\,,  \label{sr5}
\end{equation}
where
\begin{equation}
I_{0}=\frac{\omega _{1}^{2}}{(4\pi )^{4}c^{8}\hbar ^{2}r_{0}^{2}r_{1}^{2}}%
\,,\quad I_{N}=|N_{fi}(\mathbf{k}_{1},-\mathbf{k}_{0})|^{2}\,.  \label{sr6}
\end{equation}
In measurements of nuclear inelastic scattering, the signal is
accumulated over many pulses of synchrotron radiation and resulted from
scattering by many nuclei. Therefore the above expression must be summed
over the final states (index $f$) and averaged over the initial states
(index $i$) of the phonon system. We will denote such intensity as
$\overline{I(\omega _{0},t)}$ and call the average intensity.

\subsection*{4.1 Energy dependence of scattered radiation.}

It is instructive to analyze first the complete energy spectrum of the
scattered radiation when the measurement is performed during the total
delay time interval from zero to infinity and the integral intensity is
\begin{equation}
I_{int}(\omega _{0})=\int_{0}^{\infty }I(\omega _{0},t)dt\,.  \label{ii0}
\end{equation}
Since $I(t)=0$ at $t<0,$ the integration limits can be formally
replaced by $(-\infty ,\,\infty )$. Taking into account that
expression (\ref{sr5}) is the modulo squared of the Fourier integral over
frequency, we use the Parseval's theorem and obtain straightforwardly
\begin{equation}
I_{int}(\omega _{0})=I_{0}I_{N}\int \frac{d\omega }{2\pi }\left| \tilde{L%
}_{fi}(\mathbf{k}_{1},-\mathbf{k}_{0},\omega _{0}-\omega _{r}+\omega
+i\gamma )\right| ^{2}I_{M}(\omega )\,.  \label{ii2}
\end{equation}
Thus, the time-integrated intensity of scattered radiation as a
function of the middle frequency $\omega_{0}$ of incident radiation
is a convolution of the energy spectrum of inelastic scattering for the
case of monochromatic incident wave with the instrumental function as
the energy spectrum of the monochromator
$I_{M}(\omega )=\left| P(\omega )\right| ^{2}$.

In order to calculate the averaged intensity
$\overline{I_{int}(\omega _{0})}$ we have to average the function
$\left|\tilde{L}_{fi}(\mathbf{k}_{1},-\mathbf{k}_{0},\omega +i\gamma )
\right|^{2}$. Substituting (\ref{a9}) in (\ref{lpi}), we obtain
\begin{equation}
\tilde{L}_{fi}(\mathbf{k}_{1},-\mathbf{k}_{0},\omega +i\gamma )=-\sum_{m}%
\frac{\langle \chi _{f}\,|e^{-i\mathbf{k}_{1}\mathbf{r}_{n}}|\chi
_{m}\/\rangle \langle \chi _{m}\,|e^{i\mathbf{k}_{0}\mathbf{r}_{n}}|\chi
_{i}\/\rangle }{(\omega -\omega _{mi}^{(p)}+i\gamma )}\,.  \label{ii3}
\end{equation}
The square modulus of the right-hand side must be summarized over the
final states that leads to the expression
\begin{equation}
\sum_{f,m,m^{\prime }}
\frac{\langle \chi _{f}\,|e^{-i\mathbf{k}_{1}\mathbf{r}%
_{n}}|\chi _{m}\/\rangle \langle \chi _{m}\,|e^{i\mathbf{k}_{0}\mathbf{r}%
_{n}}|\chi _{i}\/\rangle }{(\omega -\omega _{mi}^{(p)}+i\gamma )}
\,\frac{\langle
\chi _{i}\,|e^{-i\mathbf{k}_{0}\mathbf{r}_{n}}|\chi _{m^{\prime
}}\/\rangle \langle \chi _{m^{\prime }}\,|e^{i\mathbf{k}_{1}\mathbf{r}%
_{n}}|\chi _{f}\/\rangle }{(\omega -\omega _{m^{\prime }i}^{(p)}-i\gamma )}
\,. \label{ii4}
\end{equation}
Here the sum over $f$ can be calculated independently
\begin{equation}
\sum_{f}\langle \chi _{m^{\prime }}\,|e^{i\mathbf{k}_{1}\mathbf{r}_{n}}|\chi
_{f}\/\rangle \langle \chi _{f}\,|e^{-i\mathbf{k}_{1}\mathbf{r}_{n}}|\chi
_{m}\/\rangle =\langle \chi _{m^{\prime }}\,||\chi _{m}\/\rangle =\delta
_{mm^{\prime }}\,.  \label{ii5}
\end{equation}
Then the dependence on the direction of the scattered wave
vanishes and, instead of the triple sum, we obtain only a sum over $m$ .

Replacing index $m\rightarrow f$, we may present the averaged
intensity in the form
\begin{equation}
\overline{I_{int}(\omega _{0})}=I_{0}I_{N}\int \frac{d\omega }{2\pi }%
I_{SS}(\omega _{0}-\omega _{r}+\omega )\,I_{M}(\omega )\,,  \label{conv}
\end{equation}
where
\begin{equation}
I_{SS}(\omega )=\langle \sum_{f}\left| \tilde{L}_{fi}(\mathbf{k}_{1},-%
\mathbf{k}_{0},\omega +i\gamma )\right| ^{2}\rangle _{T}=\sum_{i,f}\frac{%
g_{i}\,|\langle \chi _{f}\/|\,\exp (i\mathbf{k}_{0}\mathbf{r}_{n})|\chi
_{i}\/\rangle |^{2}}{(\omega _{fi}^{(p)}-\omega )^{2}+\gamma ^{2}}\,.
\label{ii6}
\end{equation}
Here $\langle \cdots \rangle _{T}$ means the thermal averaging over the
initial states, and $g_{i}$ is the weight of $i$-th
initial state of the phonon system.

Equation (\ref{conv}) is a standard recipe of taking into account the
instrumental function (see, for example, Refs. \cite{KCR,KC}).
Eq. (\ref{ii6}) for the integral intensity of \textit{scattered}
radiation coincides with the known expression for cross-section
of nuclear resonant \textit{absorption}, used by Singwi and
Sj\"{o}lander (see eq.(1) in Ref. \cite{SS}). An important
property of this formula is that absorption or emission
of phonons takes place simultaneously with resonant absorption of
incident photon by nucleus, i.e., at the instant of nuclear
excitation. The intensity of scattered radiation depends only on the wave
vector $\mathbf{k}_{0}$, whereas the wave vector $\mathbf{k}_{1}$ does
not influence the energy spectrum. As follows from eq.(\ref{ii5}), this
feature is a logical sequence of the procedure of summation over all the
possible final states of the phonon system. This treatment corresponds
to measurement of scattered quanta independently of their energy and
directions of propagation. If an experimental setup would contain
additional elements allowing selection of scattered radiation with a
particular energy, or if scattered radiation would be accepted within
essentially small solid angle, we have to use instead of eq.(\ref{sr4})
some more complicated expressions with an explicit integration over
frequency $\omega _{1}$. Besides that, we would have to use the theory
of coherent scattering, i.e., to take into account interference between
various nuclei. In general, energy dependence of radiation scattered
into some particular directions may then differ from the averaged value.

Thus, under the conditions of entire averaging over angle and
energy of scattered radiation, the energy dependence of scattered
radiation coincide entirely with the energy dependence of nuclear
inelastic absorption. This conclusion agrees with the experimental
results of Ref. \cite{CBR}. One of the reason of this coincidence is that
spontaneous emission of photon during de-excitation of a single nucleus
does not depend on the resonance condition of preceding nuclear
excitation. In particular, it does not depend on the phonon state, which
has determined conditions of nuclear excitation. That is why, the
energy spectrum of incoherent nuclear inelastic scattering does not
differ from those of nuclear inelastic absorption.
At the first step, a nucleus is excited by an incident quantum with some
energy transfer to absorption or emission of phonons. The following
evolution of excited nucleus may vary, but it does not influence the
dependence of recorded radiation on the energy of incident radiation.

Following Ref. \cite{SS}, it is convenient to represent the
function $I_{SS}(\omega )$ as integral over time. Below we
use slightly different, than in Ref. \cite{SS}, way of derivation which
is more useful for us.
Let us write the expression (\ref{ii6}) as follows
\begin{equation}
I_{SS}(\omega )=\sum_{i,f}g_{i}\,|\langle \chi _{f}\/|\,\exp
(i\mathbf{k}_{0}\mathbf{r}_{n})|\chi _{i}\/\rangle |^{2}\tilde{R}(\omega
_{fi}^{(p)}-\omega )\,,  \label{ii7}
\end{equation}
where
\begin{equation}
\tilde{R}(\omega )=\frac{1}{(\omega ^{2}+\gamma ^{2})}=\int dt\exp
(i\omega t)\,R(t)\,.  \label{ii8}
\end{equation}
Substituting the Fourier integral (\ref{ii8}) in eq.(\ref{ii7}) we obtain
\begin{equation}
I_{SS}(\omega )=\int dt\exp (-i\omega t)\,F(\mathbf{k}_{0},t)\,R(t)\,,
\label{ii9}
\end{equation}
where the time correlation function $F(\mathbf{k}_{0},t)$ depends on the
lattice vibrations and is determined by
\begin{eqnarray}
F(\mathbf{k}_{0},t) &=&\sum_{i}g_{i}\sum_{f}\langle \chi _{i}\/|\,\exp (-i%
\mathbf{k}_{0}\mathbf{r}_{n})|\chi _{f}\/\rangle \langle \chi _{f}\/|\,\exp
(i\mathbf{k}_{0}\mathbf{r}_{n})|\chi _{i}\/\rangle \exp (i\omega
_{fi}^{(p)}t)  \nonumber \\
&=&\sum_{i}g_{i}\langle \chi _{i}\/|\,\exp (-i\mathbf{k}_{0}\mathbf{r}%
_{n}(0))\,\exp (i\mathbf{k}_{0}\mathbf{r}_{n}(t))|\chi _{i}\/\rangle
\nonumber \\
&=&\langle \exp (-i\mathbf{k}_{0}\mathbf{r}_{n}(0))\,\exp (i\mathbf{k}_{0}%
\mathbf{r}_{n}(t))\rangle _{T}\,.  \label{ii10}
\end{eqnarray}
The function $R(t)$ may be calculated in explicit form
\begin{equation}
R(t)=\frac{\exp (-\gamma |t|)}{2\gamma }\,.  \label{ii11}
\end{equation}
As a result, we arrive to expression
\begin{equation}
\overline{I_{int}(\omega _{0})}=I_{0}I_{N}\int dt\exp (-i[\omega _{0}-
\omega_{r}]t)F(\mathbf{k}_{0},t)\,R(t)\,\tilde{I}_{M}(t)\,,  \label{ii12}
\end{equation}
where
\begin{equation}
\tilde{I}_{M}(t)=\int \frac{d\omega }{2\pi }\exp (-i\omega
t)I_{M}(\omega )\,.  \label{ii13}
\end{equation}

In Refs. \cite{KCR,KC} similar expression was used in order to obtain
phonon density of states from energy dependence of nuclear inelastic
absorption. Our results show that, if incoherent nuclear inelastic
scattering is ideally integrated over time, its energy dependence is
identical to that of nuclear inelastic absorption. Strictly speaking, an
integration over infinite time interval cannot be reached in real
experiments, because during a short period after synchrotron radiation
flash an acquisition system is gated against strong electronic scattering.
In order to analyze possible influence of this gating on energy spectra
of inelastic scattering, we discuss time dependence of scattered radiation
intensity.

\subsection*{4.2 Time dependence of scattered radiation.}

Eq. (\ref{sr5}) for intensity of scattered radiation can be presented as
follows
\begin{equation}
I(\omega _{0},t)=I_{0}I_{N}\,\left| Q_{fi}(\omega _{0},t)\right| ^{2}\,,
\end{equation}
where
\begin{equation}
Q_{fi}(\omega _{0},t)=\int \frac{d\omega }{2\pi }\exp (-i\omega t)\tilde{%
L}_{fi}(\mathbf{k}_{1},-\mathbf{k}_{0},\omega _{0}-\omega _{r}+\omega
+i\gamma )P(\omega )\,.  \label{dt2}
\end{equation}
Substituting (\ref{lpi}) we write this expression in the form
\begin{equation}
Q_{fi}(\omega _{0},t)=\int_{0}^{t}dt^{\prime }\exp (i[\omega _{0}-\omega
_{r}+i\gamma ]t^{\prime })L_{fi}(\mathbf{k}_{1},-\mathbf{k}_{0},t^{\prime })%
\tilde{P}(t-t^{\prime })\,.  \label{dt3}
\end{equation}
Here we use eq.(\ref{s6}) and that $\tilde{P}(t)=0$
for $t<0$. As follows from eq. (\ref{dt3}), the intensity vanishes
at $t=0$. Making a replacement of the integration variable
$t^{\prime }\rightarrow t-t^{\prime }$ and using eq. (\ref{a9})
we obtain
\begin{eqnarray}
Q_{fi}(\omega _{0},t) &=&e^{-\gamma t}e^{i(\omega _{0}-\omega
_{r})t}\sum_{m}\langle \chi _{f}\,|e^{-i\mathbf{k}_{1}\mathbf{r}_{n}}|\chi
_{m}\/\rangle \langle \chi _{m}\,|e^{i\mathbf{k}_{0}\mathbf{r}_{n}}|\chi
_{i}\/\rangle e^{-i\omega _{mi}^{(p)}t}  \nonumber \\
&&\times \int_{0}^{t}dt^{\prime }\exp (i[\omega _{r}+\omega _{mi}^{(p)}
-\omega_{0}]t^{\prime })\,e^{\gamma t^{\prime }}\,\tilde{P}(t^{\prime })\,.
\label{dt4}
\end{eqnarray}
General evaluation of this expression requires an explicit form of the
response function of the monochromator. We will consider the case where
the energy width of the instrumental function is much larger than that of
nuclear resonance. Therefore, the response function $\tilde{P}(t)$
decreases significantly on the characteristic time $t_{p}$, which is much
less than the life time of excited nuclear state $t_{0}=2/\gamma $.
Then the function $\exp(\gamma t^{\prime })$ under the integral
can be approximated by unity. It is evident that within the time interval
$0<t<t_{p}$ the integral in eq. (\ref{dt4}) increases with time
and it becomes constant when $t\gg t_{p}$. In this latter region we can
replace the upper limit by $\infty $. Omitting also a constant phase
factor which does not influence the intensity, we obtain
\begin{equation}
\hspace{-5mm}
Q_{fi}(\omega _{0},t)=e^{-\gamma t}\sum_{m}\langle \chi _{f}\,|e^{-i\mathbf{k}%
_{1}\mathbf{r}_{n}}|\chi _{m}\/\rangle \langle \chi _{m}\,|e^{i\mathbf{k}%
_{0}\mathbf{r}_{n}}|\chi _{i}\/\rangle e^{-i\omega _{mi}^{(p)}t}P(\omega
_{r}+\omega _{mi}^{(p)}-\omega _{0})\,.  \label{dt5}
\end{equation}
Repeating derivation as discussed above, we arrive to the following
expression for the intensity summarized over the final states and
averaged over the initial states
\begin{equation}
\overline{I(\omega _{0},t)}=I_{0}\exp
(-t/t_{0})I_{N}\sum_{i,f}g_{i}\,|\langle \chi _{f}\/|\,\exp (i\mathbf{k}_{0}%
\mathbf{r}_{n})|\chi _{i}\/\rangle |^{2}I_{M}(\omega _{r}+\omega
_{fi}^{(p)}-\omega _{0})\,.  \label{dt6}
\end{equation}
We obtain that within the time region $t\gg t_{p}$
the intensity decreases in an expected way, i.e. exponentially
with the life time of the excited nuclear state.

Eq. (\ref{dt6}) is similar to eq. (\ref{ii7}), derived above for the
case of nuclear inelastic absorption. Therefore, following the same
approach, we obtain
\begin{equation}
\overline{I(\omega _{0},t)}=I_{0}\exp (-t/t_{0})I_{N}\int dt^{\prime }\exp
(-i[\omega _{0}-\omega _{r}]t^{\prime })F(\mathbf{k}_{0},t^{\prime })%
\tilde{I}_{M}(t^{\prime })\,.  \label{dt7}
\end{equation}
If the life time of excited nuclear state is, as we assume here, much
longer that the duration of the flash of incident radiation provided by
the monochromator, we can neglect time dependence of the function
$R(t)$ in eq. (\ref{ii12}) and replace $R(t)$ by $R(0)$.
Then the integrated over time intensity (\ref{dt7}) coincides with
the approximate expression (\ref{ii12}), obtained by other way.

Time dependence of $R(t)$ in eq.(\ref{ii12}) allows one to study
correct behaviour of the delayed radiation for very small delay time
inside the flash provided by monochromator. Within this time interval
the intensity of scattered radiation increases from zero to a certain
maximal value. When the flash is over, the major effect is the
delayed re-emission of excited nuclear state which decays exponentially
with the time.

\section*{5. Effect of hyperfine structure of nuclear levels.}

In this section we consider more complicated case, where nuclear
levels of both ground and excited states are split by hyperfine
interaction. In eqs. (\ref{n7}), (\ref{n10}) and further down we have to
replace $\omega _{i}$ by $\omega _{i}+\Delta \omega _{\alpha }^{(i)}$,
$\omega _{j}$ by $\omega _{j}+\Delta \omega _{\beta }^{(j)}$,
$\omega _{f}$ by $\omega _{f}+\Delta \omega_{\gamma }^{(f)}$,
where indexes $\alpha ,$ $\beta $\ and $\gamma $ denote various
sublevels of hyperfine structure for the initial, intermediate, and
final states, respectively. Accordingly, nuclear wave
functions have indexes $|\phi _{i\alpha }\rangle $ and so on.
Eq. (\ref{n10}) for the scattering matrix becomes
\begin{eqnarray}
M_{fi}(\mathbf{r},t;\mathbf{r}^{\prime },t^{\prime }) &=&\frac{i\exp
(i\omega _{f\gamma ,i\alpha }^{(n)}t)}{\hbar c}\int \frac{d\mathbf{k}d%
\mathbf{k}^{\prime }}{(2\pi )^{6}}e^{i\mathbf{kr+}i\mathbf{k}^{\prime }%
\mathbf{r}^{\prime }}N_{fi}^{\gamma \alpha }(\mathbf{k},\mathbf{k}^{\prime
},t-t^{\prime })  \nonumber \\
&&\times \exp [i(-\omega _{ji}^{(n)}+i\gamma _{j})(t-t^{\prime })]e^{-i%
\mathbf{kr}_{n}(t)}e^{-i\mathbf{k}^{\prime }\mathbf{r}_{n}(t^{\prime })}\,,
\end{eqnarray}
where $\omega _{f\gamma ,i\alpha
}=\omega _{f}+\Delta \omega _{\gamma }^{(f)}-\omega _{i}-\Delta \omega
_{\alpha }^{(i)}$, and the nuclear function
$N_{fi}^{\gamma \alpha }(\mathbf{k},\mathbf{k}^{\prime },t)$
acquire the time dependence
\begin{equation}
N_{fi}^{\gamma \alpha }(\mathbf{k},\mathbf{k}^{\prime },t)=\sum_{\beta
}\langle \phi _{f\gamma }\/|\,\mathbf{e}_{f}\hat{\mathbf{j}}_{n}(%
\mathbf{k})|\phi _{j\beta }\rangle \langle \phi _{j\beta }\/|\,%
\mathbf{e}_{i}\hat{\mathbf{j}}_{n}(\mathbf{k}^{\prime })|\phi _{i\alpha
}\rangle \exp \left( i\Delta \omega _{j\beta ,i\alpha }\,t\right) \,,
\end{equation}
where $\Delta \omega _{j\beta
,i\alpha }=\Delta \omega _{\beta }^{(j)}-\Delta \omega _{\alpha }^{(i)}$.

Repeating the derivation discussed above, it is easy to obtain the
general expression for the time dependence of intensity as
\begin{equation}
I(\omega _{0},t)=I_{0}\,\sum_{\alpha ,\gamma }g_{\alpha }\left| \tilde{Q}%
_{fi}^{\gamma \alpha }(\omega _{0},t)\right| ^{2}\,,
\end{equation}
where $g_{\alpha }$ is the statistical weight of the state
$(i\alpha )$, and
\begin{eqnarray}
\tilde{Q}_{fi}^{\gamma \alpha }(\omega _{0},t) &=&\int_{0}^{t}dt^{\prime
}\exp (i[\omega _{0}-\omega _{r}+i\gamma ]t^{\prime })\,N_{fi}^{\gamma
\alpha }(\mathbf{k}_{1},-\mathbf{k}_{0},t^{\prime }) \nonumber \\
&&\times L_{fi}(\mathbf{k}_{1},-\mathbf{k}_{0},t^{\prime })\tilde{P}%
(t-t^{\prime })\,.
\end{eqnarray}
We note that hyperfine splitting can significantly exceed the width of
the nuclear resonance. Nevertheless, the characteristic time
$t_{N}\approx 1/\Delta \omega $ of nuclear function variation is much
larger than the characteristic time $t_{p}$ of the response function of
the monochromator. Therefore at $t\gg t_{p}$ we can replace
the lower limit of integration by $-\infty $ and the function,
$\exp (-\gamma t^{\prime})\,N_{fi}^{\gamma \alpha }
(\mathbf{k}_{1},-\mathbf{k}_{0},t^{\prime })$
by its value at the upper limit $t$.

As a result, instead of (\ref{dt7}), we obtain the generalized equation
\begin{equation}
\overline{I(\omega _{0},t)}=I_{0}\exp (-t/t_{0})I_{N}(t)\int dt^{\prime
}\exp (-i[\omega _{0}-\omega _{r}]t^{\prime })F(\mathbf{k}_{0},t^{\prime })%
\tilde{I}_{M}(t^{\prime })\,,
\end{equation}
where
\begin{equation}
I_{N}(t)=\sum_{\alpha ,\gamma }g_{\alpha }\left| N_{fi}^{\gamma \alpha }(%
\mathbf{k}_{1},-\mathbf{k}_{0},t)\right| ^{2}\,.
\end{equation}
To obtain the intensity $I_{N}(t)$ integrated over the angle of exit of
scattered photons, it is necessary to integrate over directions of the
vector $\mathbf{k}_{1}$. The function $I_{N}(t)$ does not depend on the
resonance condition, in particular, on the value
$(\omega _{0}-\omega _{r})$.
Therefore it does not influence the energy spectrum of incoherent
nuclear inelastic scattering.

\section*{6. Conclusion}

We present the theoretical analysis of energy and time
dependence of incoherent nuclear inelastic resonant
scattering of synchrotron radiation (SR) accompanied by absorption and
emission of phonons in crystal lattice. The theory is based on the
Maxwell's equations and the time dependent perturbation theory of the
quantum mechanics. This allows us to analyze explicitly physical nature
of the considered processes.

In order to treat the case of incoherent scattering, we assume
that, though the scattering process is coherent in general, the
conditions of measurements do not allow experimental observation of the
interference term in the intensity of scattered radiation due to its
smearing out during averaging over various parameters. This concept of
incoherent scattering significantly simplify calculations, because it
avoids calculations of the interference term which would in any event
vanish during averaging. Our results show that if measurements are
performed without energy or angular analysis of scattered radiation, the
dependence of its intensity on the energy of incident radiation is
identical to that of nuclear inelastic absorption.
Formally, this is caused by averaging of scattering intensity over final
states of phonon system.

The measured spectra are influenced only by processes of absorption and
emission of phonons only during the nuclear excitation by the SR flash.
We assume that x rays emitted by various electrons in a storage ring
are incoherent, and radiation from single electron can be considered as
instantaneous. This assumption is based on the experimental conditions,
where the intensity of scattered radiation is accumulated after many
SR flashes, whereas only one photon is recorded after each SR flash.
We show that the possibility to measure the density of phonon states
under these conditions arises from the property of monochromator to delay
the initially instantaneous SR flash up to the time interval which
significantly exceeds the period of nuclear vibrations.
It is sufficient to describe the case of incoherent scattering within
the first Born approximation. Then the calculation is reduced to
thermal averaging of the phase factor depending on the nuclear
coordinates at the instant of photon both absorption and emission.
Within this approach, the time correlation function can be calculated
with a proper account for multi-phonon processes,
whereas treating the multi-phonon processes in the case of coherent
scattering is significantly more complicated problem.\\[3mm]
\centerline{*\qquad *\qquad *}
The work is supported by the Russian Foundation for the Basic Research
(project N.\,01-02-16508).

\end{document}